\documentclass[aps]{revtex4}
\usepackage{epsfig}
\usepackage{amsmath,amssymb,color}
\usepackage{epstopdf,epsfig}
\parskip=\medskipamount

\newcommand{\eq}[1]{(\ref{#1})}
\newcommand{\fig}[1]{Fig.\ref{#1}}

\newcommand{\be}{\begin{equation}}
\newcommand{\ee}{\end{equation}}

\newcommand\disp{\displaystyle}

\newcommand{\la}{\left<}
\newcommand{\ra}{\right>}

\begin{document}

\title{Path counting on simple graphs: from escape to localization}

\author{S.K. Nechaev$^{1}$, M.V. Tamm$^{2,3}$, and O.V. Valba$^{3,4}$}
\address{$^{1}$J.-V. Poncelet Laboratory, CNRS, UMI 2615, 11 Bolshoy Vlasievski,
119002 Moscow, Russia, \\ $^{2}$Physics Department, Moscow State University, 119992, Moscow,
Russia, \\ $^{3}$Department of Applied Mathematics, National Research University Higher School of
Economics, 101000, Moscow, Russia, \\ $^{4}$N.N. Semenov Institute of Chemical Physics RAS, 119991,
Moscow, Russia.}

\begin{abstract}

We study the asymptotic behavior of the number of paths of length $N$ on several classes of
infinite graphs with a single special vertex. This vertex can work as an entropic trap for the
path, i.e. under certain conditions the dominant part of long paths become localized in the
vicinity of the special point instead of spreading to infinity. We study the conditions for such
localization on decorated star graphs, regular trees and regular hyperbolic graphs as a function
of
the functionality of the special vertex. In all cases the localization occurs for large enough
functionality. The particular value of transition point depends on the large-scale topology of
the
graph. The emergence of localization is supported by the analysis of the spectra of the adjacency
matrices of corresponding finite graphs.
\end{abstract}

\maketitle

\section{Introduction}

In this work we study the asymptotic behavior of the total number of paths of length $N$ on several
classes of regular graphs. We call this a "path counting" (PC) problem as opposed to a more usual
"random walk" problem (RW) which studies the distribution of the end points of symmetric random
walks on graphs. The difference between PC and RW is in different normalizations of the elementary
step: for PCs all steps enter in the partition function with the weight one, while for symmetric
RWs, the step probability depends on the vertex degree, $p$: the probability to move along each
graph bond equals $p^{-1}$. For graphs with a fixed vertex degree, the PC partition function and
the RW probability distribution differ only by the global normalization constant, and corresponding
averages are indistinguishable. However, for inhomogeneous graphs the distinction between PC and RW
is crucially: in the path counting problem "entropic" localization of the paths may occur, while it
never happens for random walks. The distinction between PC and RW, and the entropic localization
phenomenon were first reported for self-similar structures in \cite{17} and later were rediscovered
for star graphs in \cite{ternovsky}.
More recently this phenomenon was studied for regular lattices with defects in \cite{burda} where
authors introduce a notion of ``maximal entropy random walk'' which is essentially identical to our
path-counting problem.

\begin{figure}[ht]
\epsfig{file=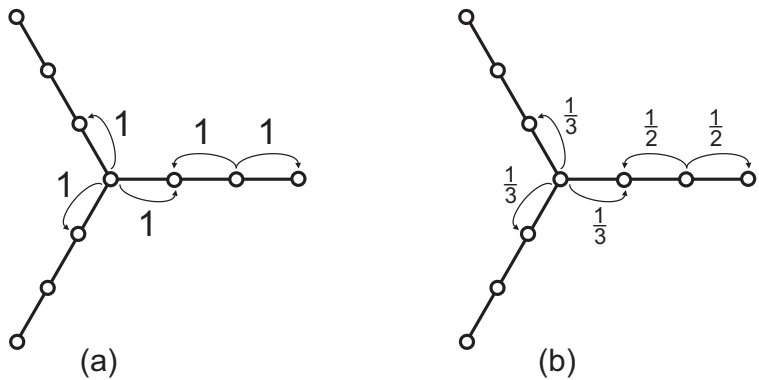, width=10cm}
\caption{Star-like graph with $p=3$ branches. In a) we enumerate trajectories and each step
carries a weight 1. In b) the random-walk the local transition probabilities satisfy conservation
condition in each vertex.}
\label{fig:01}
\end{figure}

Following \cite{ternovsky} we begin with star-like discrete graphs, ${\cal G}$, i.e. a union of $p$
discrete half-lines joint together in one point (the root) -- see the \fig{fig:01}a -- just this
model was a subject of the work \cite{ternovsky}. Regard all $N$-step discrete trajectories on
${\cal G}$, and define $Z_N(x)$, the total number of trajectories starting from the root of the
graph, and ending at distance $x = 0, 1, 2,...$ from the root (regardless on which particular
branch the path ends). One can consider $Z_N(x)$ as a partition function of an ideal polymer chain
with $N$ links with one end fixed in the root of the graph ${\cal G}$ and another end to be
anywhere. Given the partition function $Z_N(x)$, one can define the corresponding averages:
\be
\la x^2(N|p) \ra = \frac{\disp \sum_{x=0}^{\infty} x^2 Z_N(x|p)}{\disp \sum_{x=0}^{\infty}
Z_N(x|p)}.
\label{eq:01}
\ee
The straightforward computations \cite{ternovsky} show that the asymptotic behavior of $\la
x^2(N|p) \ra$ at $N\to \infty$ for large $N$ depends drastically on the number of branches, $p$, in
the star-like graph ${\cal  G}$. Indeed,
\be
\la x^2(N|p) \ra_{N\to \infty}\to f(p) \times \left\{\begin{array}{ll} O(N) , & \quad p=1,2;
\medskip \\ O(1), & \quad p=3,4,... \end{array} \right.
\label{eq:02}
\ee
where $f(p)$ is some positive function which depends on $p$ only and does not depend on $N$. In
other words, for $p=1,2$ (the half-line and the full line) the trajectories on average diverge from
the origin with the typical distance proportional to $\sqrt{N}$, as one would naturally expect for
a regular random walk. Besides, for $p>2$ the trajectories on average stay localized in the
vicinity of the junction point.

To understand qualitatively this behavior, note that the recursion relation connecting $Z_N(x|p)$
with $Z_{N+1}(x|p)$ crucially depends on whether $x$ is the root point ($x=0$), or not. Indeed, for
each trajectory of length $N$ ending at a point $x \ge 1$ there are exactly two possible ways to
add $(N+1)$-th step to it, thus each such path "gives birth" to two paths of length $(N+1)$.
Contrary to that, for each $N$-step path which ends at $x=0$, there are $p$ different ways of
adding a new step. Therefore, for $p>2$ passing to $x=0$ becomes entropically favorable, and the
root point plays a role of an effective "entropic trap" for trajectories.

Let us emphasize that this peculiar behavior of the partition function (as a function of $p$) is
specific to the \emph{path counting} problem, and manifests itself in the \emph{equilibrium}
(combinatoric) computations of ideal polymer conformational statistics. Contrary to that, one can
think of a closely related \emph{non-equilibrium} problem, namely calculation of a probability
distribution, $P_N(x|p)$, for the end-to-end distance of $N$-step random walk on the star graph of
$p$ branches. In that case, the probability distribution, due to the normalization condition,
should be integrated into 1 on each step, which leads to the obvious normalization of $P_N(x,p)$:
\be
\sum_{x=0}^{\infty} P_N(x|p) = 1 \; \mbox{for any}\; N.
\label{eq:03}
\ee
Therefore, the entropic advantage to stay at the origin is compensated by the fact that possible
steps from the origin have probability $p^{-1}$ instead of $1/2$. In other words, if in the path
counting problem all trajectories have equal weights 1, in the random walk problem the trajectories
have weights $2^{-(N-n)}p^{-n}$, where $n$ is the number of returns to the point $x=0$, varying
from a path to path. It is easy to see that in the random walk problem
\be
\lim_{N\to\infty} \frac{\la x^2(N|p) \ra}{N} =\lim_{N\to\infty} \frac{1}{N} \sum_{x=0}^{\infty} x^2
P_N(x|p) = {\rm const} > 0
\label{eq:03a}
\ee
regardless the value of $p$.

The qualitative arguments supported by exact computations for specific models, demonstrate that
entropic localization in the path counting occurs in inhomogeneous systems with broken
translational invariance. On uniform trees there is no entropically favorable vertices and PC does
not exhibit any localization transitions \cite{texier}. However, as we see below, the localization
is topology-dependent phenomena and occurs on decorated graphs.

The paper is organized as follows. In the Section II we consider finite tree-like regular graphs
with a special vertex (``entropic trap'') at the origin, and compute the asymptotics of its
partition function based on the spectral properties of the graph adjacency matrix. We show,
however, that this approach has some limitations: even increasing the size of the tree to infinity
we cannot capture a non-localized solution properly. Indeed, it is not surprising: a finite tree of
any size has a non-vanishing fraction of nodes with the degree "1" (terminal ``leaves''). In turn,
an infinitely large tree does not have such vertices, and therefore not all of its properties can
be recovered by studying the sequence of increasing finite graphs. To resolve this problem in the
Section III (which plays the central role in the paper), we study infinite tree-like graphs with a
special vertex, and show that depending on the functionality of the vertex, there indeed exists a
transition between localized and delocalized state. In the Section IV we generalize the results for
two other families of graphs with a special entropically attractive point (we call these two
families ``decorated star graphs'' and ``regular hyperbolic graphs''). In the last Section we
summarize and discuss the obtained results and formulate some open questions.

\section{Path counting on finite tree-like graphs}

Consider an arbitrary graph $\cal{G}$ with the adjacency matrix $B_{\cal{G}}$. It is easy to see
that the partition function $Z_N$ described above can be easily expressed in terms of
$B_{\cal{G}}$. Indeed, matrix elements of $B_{\cal{G}}^N$,
\be
\la i | B_G^N | j \ra
\ee
enumerate walks of length $N$ starting in vertex $j$ and ending in vertex $i$. Therefore, e.g., the
total number of paths starting at $i$-th vertex and ending at distance $x$ from it, $Z^{(i)}_N(x)$,
equals the sum of the matrix elements over all $j$ with a given distance $x$ from $i$
\be
Z^{(i)}_{N}(x)= \sum_{j: \text{dist} (i,j) = x}\la i | B_G^N | j \ra.
\ee
Clearly, this means that asymptotic behavior of $Z_{N}$ is controlled by the largest eigenvalue of
$B_{\cal G}$, $\lambda_\text{max} ({\cal G})$. More precisely, in the large $N$ limit
\be
\log Z_{N}(x)\approx N \log \lambda_\text{max} ({\cal G}) + o(N).
\label{groundstate}
\ee
Note that for bimodal graphs there always exists a symmetrical pair of largest eigenvalues $\pm
\lambda_\text{max} (\cal{G})$; as a result $Z_{N}(x)$ alternate between the value prescribed by
\eq{groundstate} for even $(N+x)$ and 0 for odd $(N+x)$. In what follows this trivial alternating
behavior will appear recurrently and will not be specially mentioned.

\begin{figure}[ht]
\epsfig{file=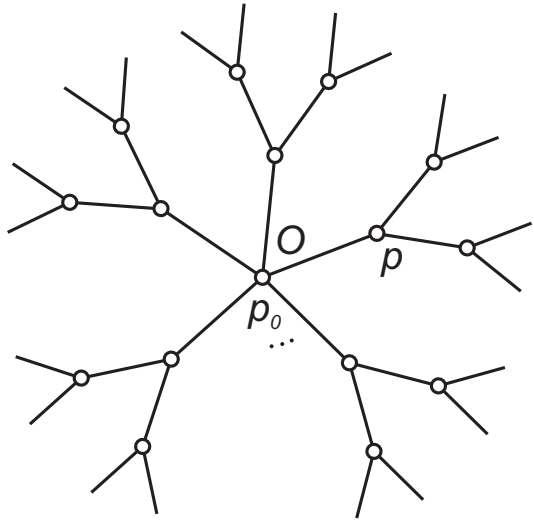,width=7cm}
\caption{Image of a regular $p$-branching Cayley tree with $p_0$ branches at the origin $x=0$.
The particular case of a tree with $p=3, n = 3$ is shown.}
\label{fig:02}
\end{figure}

\begin{figure}[ht]
\epsfig{file=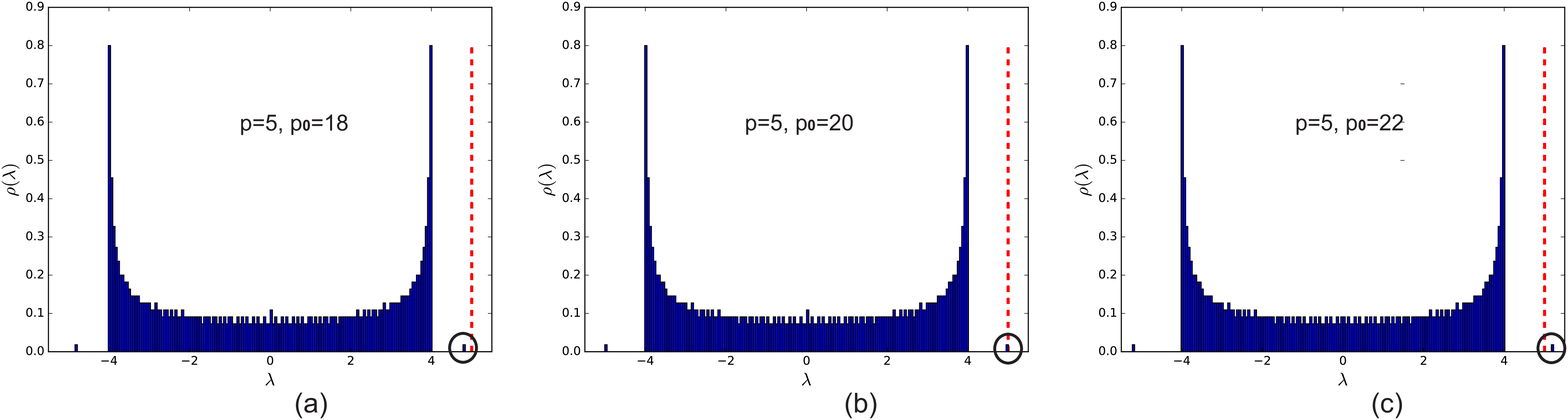, width=16cm}
\caption{Spectral density of the matrix $A_{1000}$ (a) the localization transition,
(b) at the transition point and (c) in the localized phase.  Red line indicates $\lambda=p$.}
\label{fig:spec}
\end{figure}

As a particular example of ${\cal G}$, consider a regular branching tree-like graph with $p_0$
branches coming out of the origin, and $p$ branches out of any other vertex (as we are interested
in a possible localization at the origin, we suppose $p_0\ge p$). The maximum number of generations
is $n$. The number of vertices in such a graph grows exponentially with $n$, so the direct analysis
of its spectrum might seem challenging. However, it turns out that one can drastically simplify the
problem by exploiting the symmetries of $\cal G$. Indeed, according to \cite{rojo2005,rojo2007} the
set of eigenvalues of $B_{\cal {G}}$ coincides with the set of eigenvalues of a tri-diagonal
symmetric matrix $A_n$ with elements $a_{ij}^{(n)}$, which are defined as follows:
\be
\begin{cases}
a_{i,i}^{(n)} = 0 \medskip \\
a_{n,n-1}^{(n)}=a_{n-1,n}^{(n)}=\sqrt{p_0} \medskip \\
a_{i,i-1}^{(n)}=a_{i-1,i}^{(n)} = \sqrt{p-1}; \qquad (i = 2,...,n-1).
\end{cases}
\ee
The eigenvalues of $j \times j$ submatrices correspond to multiply degenerated eigenvalues of
$B_{\cal {G}}$  (the multiplicity of eigenvalues equals the number of vertices at the generation
$n-j$ of the tree from the root point). The corresponding eigenvectors are localized on the outer
branches, exactly vanishing at the lowest $n-j$ generations, and take alternating values at the
adjacent outer sub-branhces. The eigenvalues of the whole matrix $A_n$ are non-degenerate and their
eigenvectors span the whole tree. It follows immediately, that the largest eigenvalue of $B_{\cal
{G}}$ which is of crucial importance in order to estimate the asymptotics of $Z_N$, is an
eigenvalue of the matrix $A_n$ itself: indeed, its eigenvector should be positively defined. Thus,
the study of the spectrum of an exponentially large matrix $B_{\cal {G}}$, is reduced to a similar
study of a small and simple matrix $A_n$, which can be easy treated both numerically (see the
\fig{fig:spec}) and analytically.

The characteristic polynomial $\mathrm{P}_n(p_0,p)$ for the matrix $A_n$ satisfies the recursion
relation \cite{KovalevaValbaNechaev}:
\be
\begin{cases}
\mathrm{P}_{n}(p_0,p)=\lambda \mathrm{P}_{n-1}(p_0,p) -(p-1) \mathrm{P}_{n-2}(p_0,p); \medskip \\
\mathrm{P}_{0}(p_0,p)=\frac{p_0}{p-1}; \medskip \\
\mathrm{P}_{1}(p_0,p)=\lambda.
\end{cases}
\ee
This system is easy to solve if one looks for $\mathrm{P}_n(p_0,p)$ in the form
\be
\mathrm{P}_n(p_0,p) = C_+\mu_+^{n} +C_-\mu_-^{n},
\ee
giving
\be
\disp \mu_{\pm}=\frac{{\lambda} \pm \sqrt{{\lambda}^2 - 4(p-1)}}{2},
\ee
and
\be
\disp C_{\pm}=\frac{p_0}{2(p-1)} \mp \frac{\lambda(p_0-2(p-1))}{2(p-1)\sqrt{{\lambda}^2 -
4(p-1)}}.\medskip
\ee

It is easy to see that the resulting equation $\mathrm{P}_n(p_0,p)  = 0 $ is even with respect to
$\lambda$. To solve it, define a new variable $\phi$ by
\be
\lambda= \pm 2\sqrt{p-1}\cosh{\phi},
\ee
where $|\lambda| \ge 2\sqrt{p-1}$ corresponds to real $\phi$ and $\lambda < 2\sqrt{p-1}$ -- to
purely imaginary $\phi$. Then
\be
\begin{array}{l}
\disp \mu_{\pm}=\sqrt{p-1}e^{\pm\phi}, \medskip \\
\disp C_{\pm}=\frac{p_0}{2(p-1)}\pm \frac{p_0-2(p-1)}{2(p-1)\tanh\phi},
\end{array}
\ee
and the equation $\mathrm{P}_n(p_0,p)  = 0$ becomes
\be
\tanh n \phi=\frac{p_0}{p_0-2(p-1)}\tanh \phi,
\ee
which for any $n$ has many imaginary solutions and a single real one, which corresponds to
$\lambda_{\text max}^{(n)}$. The limit of this solution for $n \to \infty$ is
\be
\lambda_{\max}= \lim_{n\to\infty} \lambda_{\max}^{(n)} = \frac{p_0}{\sqrt{p_0-p+1}}.
\label{maxLambda}
\ee

Therefore, we conclude that on any large but finite tree the number of trajectories in the $N \to
\infty$ limit behaves asymptotically as (see eq. \eq{groundstate})
\be
Z_N(x, p, p_0, n) \sim  (\lambda_{\text max})^N = \left(  \frac{p_0}{\sqrt{p_0-p+1}} \right)^N.
\label{zn_finite}
\ee
This result, however, looks a bit strange after close examination: for a partition function on an
{\it infinite} tree there exists (for $p_0 \geq p$) a lower bound
\be
Z_N(x, p, p_0, n=\infty) \geq p^N.
\label{zn_estimate}
\ee
Indeed, on each step there exist at least $p$ different directions to go, which seems to contradict
\eq{zn_finite} for $p < p_0 < \bar{p}_0=p^2-p$. This apparent discrepancy is, of course, due to the
order of taking limits $n \to \infty$ and $ N \to \infty$. In a system with large but finite $n$
there is always a finite fraction of terminal nodes (``leaves'' of the tree) with degree "1",
violating the reasoning behind \eq{zn_estimate}. In turn, in a system with infinite $n$, as we show
in the next Section, there exists an additional eigenvalue of the adjacency matrix equal to $p$,
corresponding to a density wave spreading with finite velocity from the root point to infinity.
Depending on which of two eigenvalues, the one given by \eq{maxLambda}, or this new, $\lambda = p$,
is maximal, the partition function of the infinitely large system is either localized, or,
respectively, delocalized.

\section{Localization of trajectories on an infinite tree with a "heavy" root}

Consider the same graph ${\cal G}$ as defined above (see \fig{fig:02}) but with infinitely large
number of generations $n$. We start with writing explicitly the recursion relation for the
partition function, $Z_N(x)$, of all $N$-step paths on ${\cal G}$, starting at the origin and
ending at some distance $x$ from it:
\be
\left\{\begin{array}{rcll} \disp Z_{N+1}(x) & = & (p-1)Z_N(x-1) + Z_N(x+1), & \quad x\ge 2
\medskip \\ \disp Z_{N+1}(x) & = & p_0 Z_N(x-1) + Z_N(x+1), & \quad x=1 \medskip \\
Z_{N+1}(x) & = & Z_N(x+1), & \quad x=0 \medskip \\
Z_{N}(x) & = & 0, & \quad x\le -1 \medskip \\
Z_{N=0}(x) & = & \delta_{x,0},
\end{array} \right.
\label{eq:04}
\ee
where $x$ is the distance from the root of the Cayley graph ${\cal  G}$, measured in number of
generations of the tree.

In order to solve this set of equations \cite{koleva,gangardt} we make a shift $x\to x+1$,  and
substitute
\be
Z_N(x) = A^N B^x W_N(x).
\label{eq:zw}
\ee
with $A=B=\sqrt{p-1}$. This substitution allows to symmetrize the original equation, which in terms
of $W$ takes now the form
\be
\left\{\begin{array}{rcll} W_{N+1}(x) & = & \disp W_N(x-1) + W_N(x+1) +
\frac{p_0-p+1}{p-1}\,\delta_{x,2}\,W_N(x-1), & \qquad x\ge 1 \medskip \\
W_N(x) & = & 0, & \qquad x=0 \medskip \\
W_{N=0}(x) & = & \disp \frac{\delta_{x,1}}{\sqrt{p-1}}.
\end{array} \right.
\label{eq:06}
\ee
Note that this equation can be written in the matrix form
\be
\mathbf{W}_{N+1}=T\,\mathbf{W}_N; \qquad \mathbf{W}_0=((p-1)^{-1/2},0, \dots)^{\intercal},
\ee
where the transfer matrix $T$ is an infinite tri-diagonal matrix
\be
T= \left(\begin{matrix}
0 & 1 & 0 & 0& \dots\\
\frac{p_0}{p-1} & 0 & 1 & 0&  \dots\\
0 & 1 & 0 &1&   \dots \\ 
0& 0& 1& 0 & \dots &  \\
\vdots&\vdots&\vdots& \vdots & \ddots  \\
\end{matrix}\right),
\label{TRmatrix}
\ee
whose $n$-th main minors are equal to $\sqrt{p-1} A_n$

Introducing the generating function
\be
{\cal  W}(s,x) = \sum_{N=0}^{\infty}W_N(x) s^N \qquad \left(W_N(x) = \frac{1}{2\pi i}\oint {\cal
W}(s,x) s^{-N-1}\, ds \right)
\label{eq:07}
\ee
and its $\sin$--Fourier transform
\be
\tilde{{\cal  W}}(s,q) = \sum_{x=0}^{\infty}{\cal  W}(s,x) \sin qx \qquad \left({\cal  W}(s,x) =
\frac{2}{\pi} \int_{0}^{\pi} \tilde{{\cal  W}}(s,q) \sin qx\, dq \right),
\label{eq:08}
\ee
one obtains from \eq{eq:06}
\be
\frac{\tilde{{\cal  W}}(s,q)}{s} - \frac{\sin q}{s\sqrt{p-1}} = 2\cos q\, \tilde{{\cal  W}}(s,q) +
\frac{2}{\pi} \frac{p_0-p+1}{p-1} \sin 2q \int_{0}^{\pi} \tilde{{\cal W}}(s,q) \sin q\, dq.
\label{eq:09}
\ee
Rewriting \eq{eq:09} as
\be
\tilde{{\cal W}}(s,q) = \frac{1}{\sqrt{p-1}} \frac{\sin q}{1-2s\cos q} +
\frac{2s(p_0-p+1)}{\pi(p-1)}\, \frac{\sin 2q}{1-2s\cos q}\int_{0}^{\pi} \tilde{{\cal W}}(s,q)\sin
q\, dq,
\label{eq:10}
\ee
multiplying both sides of \eq{eq:10} by $\sin q$ and integrating over $q$, $q\in [0,\pi]$ , one
arrives to an algebraic equation for
\be
I(s) = \int_0^{\pi} \tilde{{\cal W}}(s,q) \sin q \, dq,
\label{eq:11}
\ee
namely
\be
I(s) = \frac{1}{\sqrt{p-1}} \int_0^{\pi} \frac{\sin^2 q}{1-2s\cos q}\, dq + I(s)\,
\frac{2s(p_0-p+1)}{\pi(p-1)}\, \int_0^{\pi} \frac{\sin q\, \sin 2q}{1-2s \cos q}\, dq.
\label{eq:12}
\ee
The solution of this equation reads
\be
I(s) = \frac{\disp \frac{1}{\sqrt{p-1}} \int_0^{\pi} \frac{\sin^2 q}{1-2s\cos q}\, dq}{\disp
1-\frac{2s(p_0-p+1)}{\pi(p-1)} \int_0^{\pi} \frac{\sin q\, \sin 2q}{1-2s \cos q}\, dq} =
\frac{\pi\sqrt{p-1}\,\left(1-\sqrt{1-4s^2}\right)}{\disp
4s^2(p-1)-(p_0-p+1)\left(1-\sqrt{1-4s^2}\right)^2}.
\label{eq:13}
\ee
Substituting $I(s)$ into \eq{eq:10}, and performing the inverse Fourier transform, we arrive at the
following explicit expression for the generating function ${\cal W} (s,x)$:
\begin{multline}
{\cal W}(s,x)=\frac{2}{\pi}\int_0^{\pi}\tilde{{\cal W}}(s,q) \sin qx\, dq \\ =
\frac{1}{s\sqrt{p-1}}\left(\frac{1-\sqrt{1-4s^2}}{2s}\right)^x \left(1+
\frac{2(p_0-p+1)\left(1-\sqrt{1-4s^2}\right)}{4s^2(p-1)-(p_0-p+1)
\left(1-\sqrt{1-4s^2}\right)^2}\right).
\label{eq:14}
\end{multline}

Since, by definition, $Z_N(x) = A^N B^x W_N(x)$ (see \eq{eq:zw}), we can write down the relation
between the generating functions of $Z_N(x)$ and of $W_N(x)$:
\be
{\cal Z}(\sigma,x)=\sum_{N=0}^{\infty} Z_N(x) \sigma^N = \sum_{N=0}^{\infty} A^N B^x W_N(x)
\sigma^N=B^x {\cal W}(\sigma A,x).
\ee
Thus,
\be
{\cal Z}(\sigma,x|p,p_0) = (p-1)^{x/2}{\cal W}(\sigma \sqrt{p-1},x),
\label{eq:15}
\ee
where ${\cal W}(\sigma \sqrt{p-1},x)$ is given by \eq{eq:14} where we should substitute $\sigma
\sqrt{p-1}$ for $s$. Thus, the grand partition function, ${\cal Z}(\sigma,x|p,p_0)$, of the initial
path counting problem reads
\be
{\cal Z}(\sigma,x|p,p_0) = \frac{2p_0 \sigma \left(\frac{\disp 1 - \sqrt{1 - 4
\sigma^2(p-1)}}{\disp 2\sigma} \right)^x}{2\sigma^2p_0
(p-1)-(p_0-p+1)\left(1-\sqrt{1-4\sigma^2 (p-1)}\right)}.
\label{eq:16}
\ee

The partition function, $\bar{Z}(N|p,p_0)$, of all paths starting at the origin, can be obtained by
the summation over $x$:
\be
\bar{{\cal Z}}(\sigma|p,p_0) = \sum _{x=0}^\infty {\cal Z}(\sigma,x|p,p_0).
\label{eq:07a}
\ee
Straightforward computations lead us to the following result
\begin{equation}
\bar{{\cal Z}}(\sigma|p,p_0) = \frac{4p_0 \sigma^2\left(1 - \sqrt{1 - 4 \sigma^2(p-1)}
\right)}{\left[ 2\sigma - 1 + \sqrt{1-4\sigma^2 (p-1)} \right] \left[ 2\sigma^2p_0
(p-1)-(p_0-p+1)\left(1-\sqrt{1-4\sigma^2 (p-1)}\right)\right]}.
\label{eq:16a}
\end{equation}

To extract the asymptotic behavior of the partition function
\be
\bar{Z}_N(p,p_0)= \frac{1}{2\pi i}\oint \bar{Z}(\sigma|p,p_0)\sigma^{-N-1}d \sigma
\ee
as a function of $N$, one should analyze the behavior of $\bar{{\cal Z}}(\sigma|p,p_0)$ (see
Eq.\eq{eq:16a}) at its singularities. There are three of them, namely
\be
\sigma_1 = \frac{1}{2\sqrt{p-1}}
\label{eq:sing1}
\ee
for a branching point of the square root, and
\be
\sigma_2 = \frac{1}{p}, \quad \sigma_3=\frac{\sqrt{p_0-p+1}}{p_0}
\label{eq:sing2}
\ee
for zeroes of the first and second factors in the denominator of \eq{eq:16a}, respectively. The
asymptotic behavior is governed by the dominant singularity, i.e. the one with the least absolute
value. It is instrumental to compare these singularities with eigenvalues of corresponding
finite-size problem discussed in the previous section. Indeed, $\sigma_3$ is nothing but
$\lambda_{\max}^{-1}$ given by eq.\eq{maxLambda}, $\sigma_1$ corresponds to the border of
quasi-continuous spectrum shown in \fig{fig:spec}, and $\sigma_2$ is, as discussed at the end of
the previous Section, the solution which runs away from the origin, and is therefore unavailable in
the finite-system case.

Since $\sigma_1 > \sigma_2$ for any $p \geq 3$ regardless of the value of $p_0$, the
square-root singularity never dominates. In turn, $\sigma_3$ equals $\sigma_2$ at the critical
value $p_0$ defined by the equation:
\be
\bar{p}_0=p^2-p.
\label{eq:18}
\ee
For $p_0 < \bar{p}_0$ the singularity at $\sigma_2$ gives the dominant contribution to the
partition function and in the large $N$ limit the total number of paths scales as
\be
\left.\bar{Z}(N|p,p_0)\right|_{N\gg 1} \approx \sigma_2^{-N} = c(p,p_0)\, p^N,
\label{eq:scale1}
\ee
where $c(p,p_0)$ is $N$-independent. The behavior of $\bar{Z}(N\gg 1|p,p_0)$ in Eq.\eq{eq:scale1}
should be compared to the one for $p_0=p$, where $\bar{Z}(N|p,p_0=p)=p^N$ for any $N$ (for $p_0=p$
there are always exactly $p$ different ways to add a $N$th step to any $(N-1)$-step trajectory). We
see that in this regime, the root has essentially no influence on the asymptotics of a partition
function, and any typical $N$-step trajectory ends at the distance $\bar{x} = \frac{p-2}{p}N$ from
the origin. Since there are always more possibilities to go away from the origin than to go back to
it, there is a finite drift, with Gaussian fluctuations of order $\Delta x \sim \sqrt{N}$ around
the mean value of $\bar{x}$. We address the reader to \cite{texier} where the statistics of the
trajectories on regular Cayley trees is discussed in details.

\begin{figure}[ht]
\epsfig{file=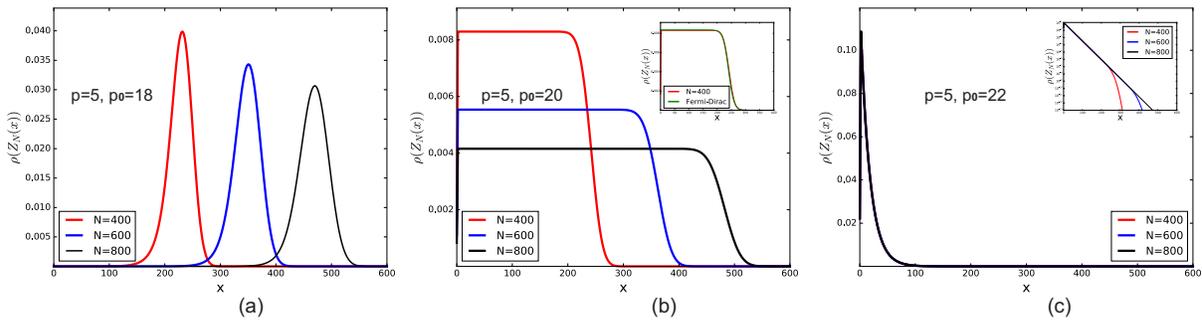, width=16cm}
\caption{The distribution of the the trajectory endpoint $\rho(Z_N(x))=Z(N,x|p,p_0)
\left[\bar{Z}(N|p,p_0)\right]^{-1}$ as a function of $x$ (a) the localization transition, (b)
at the transition point (Insert: numerical results a fitted by the Fermi-Dirac distribution with
the Fermi-energy$e_{F}=\bar{x} =\frac{p-2}{p}N$) and (c) in the localized phase (Inset: in the
logarithmic scale).}
\label{fig:03}
\end{figure}

Contrary to that, for $p_0 > \bar{p}_0$ the large-$N$ scaling of the number of paths significantly
depends on $p_0$:
\be
\bar{Z}(N|p,p_0) \approx \sigma_3^{-N} \sim \left(\frac{p_0}{\sqrt{p_0-p+1}}\right)^N,
\label{eq:scale2}
\ee
which is a signature of the localization. Indeed, the very fact that the partition function depends
on $p_0$ for any $N$ indicates that typical trajectories return to the origin for any $N$. To have
better understanding of the typical behavior of trajectories, we insert the critical value
$\sigma_3$ into \eq{eq:16}, which results in the following $x$-dependence of the partition
function
\be
Z(N,x|p,p_0) \approx \sigma_3^{-N} \sim \left(\frac{p_0}{\sqrt{p_0-p+1}}\right)^N
\left(\frac{p-1}{\sqrt{p_0-p}}\right)^x,
\label{eq:scale2x}
\ee
Eq.\eq{eq:scale2x} indicates the exponential decay of $Z(N,x|p,p_0)$ as a function of $x$. That
behavior (as well as Eq.\eq{eq:18}) is confirmed by direct iterations of Eqs.\eq{eq:04} for $p=5$
and $p_0 = 17$ (below the localization transition point), $p_0=20$ (at the transition point) and
$p_0=22$ (in the localized phase), see \fig{fig:03}. Note that exactly at the transition point
$p_0=\bar{p}_0$ the distribution of the the trajectory endpoint is nicely approximated by the
Fermi-Dirac distribution.

\section{Path counting on decorated star graphs and regular hyperbolic graphs}

\begin{figure}[ht]
\epsfig{file=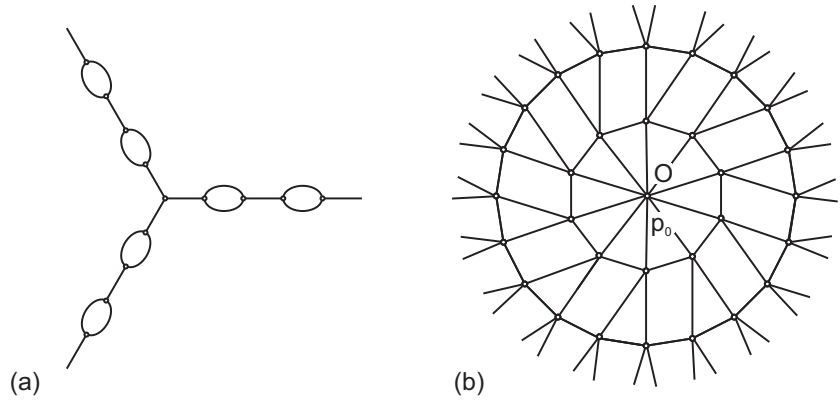, width=11cm}
\caption{Decorated star graphs: (a) Simplest decorated star graph with $p_0=p=3$; (b)
Generic decorated graph with $p_0=10, p=5, b=2$.}
\label{fig:04}
\end{figure}

Here we aim to generalize the above results for two classes of more general graphs. One class,
called a ``decorated star graph'', is shown on \fig{fig:04}a, it consists of $p_0$ bundles such
that each bundle has an overall linear topology, but all vertices in the bundle have functionality
$p$. The second class is a class of ``regular hyperbolic graphs'' with special point at the origin
(see \fig{fig:04}b). Here, once again, $p_0$ bonds are originating from the root and each vertex
except the root has functionality $p$. However, among $p$ bonds originating from a node at distance
$x$ from the root, one bond is going ``down'' to a point at distance $(x-1)$, $(p-b-1)$ are going
``up'' to points at distance $(x+1)$, and $b$ ``horizontal'' bonds connect the node with others at
the same distance $x$ from the origin.

Clearly, star graphs considered in the Introduction are decorated stars with $p=2$, while regular
trees with heavy root considered in sections II-III are regular hyperbolic graphs with $b=0$. In
this section we aim to understand how the additional parameter, $p$ in the first case, and $b$ in
the second, influences the localization transition point. Since the mathematical structure of these
two problems is extremely similar, we discuss them in parallel.

\begin{figure}[h]
\epsfig{file=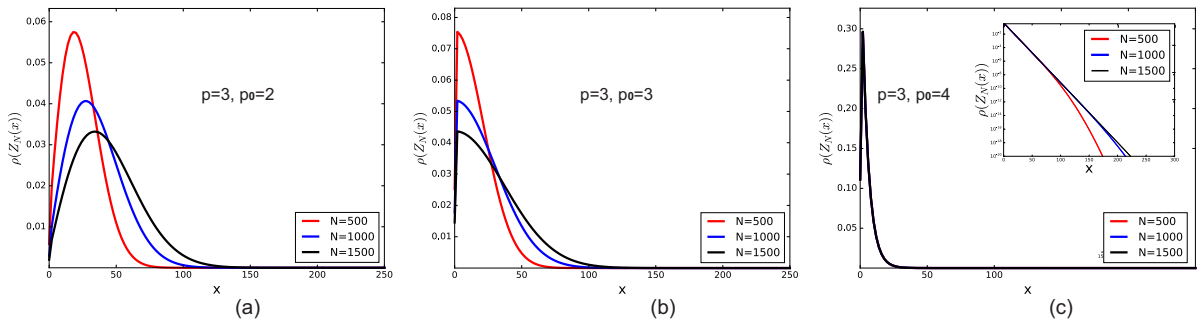, width=16cm}
\caption{The distribution of the the trajectory endpoint
$\rho(Z_N(x))=Z(N,x|p,p_0)\left[\bar{Z}(N|p,p_0)\right]^{-1}$ as a function of $x$ (a) below the
localization transition, (b) at the transition point and (c) in the localized phase (Inset: in the
logarithmic scale). }
\label{fig:05}
\end{figure}

Consider first a decorated star graph. The partition function, $Z_N(x)$, of all $N$-step paths on
such a graph, starting at the origin and ending at some distance $x$ from the root (regardless of
which particular branch it is on), satisfies the recursion (compare with \eq{eq:04})
\be
\left\{\begin{array}{rcll} \disp Z_{N+1}(x) & = & Z_N(x-1) + (p-1)Z_N(x+1), & \quad x= 2k+1 \
(k\ge1)
\medskip \\
 \disp Z_{N+1}(x) & = & (p-1)Z_N(x-1) + Z_N(x+1), & \quad x= 2k \ (k\ge1)
\medskip \\ \disp Z_{N+1}(x) & = & p_0 Z_N(x-1) + (p-1)Z_N(x+1), & \quad x=1 \medskip \\
Z_{N+1}(x) & = & Z_N(x+1), & \quad x=0 \medskip \\
Z_{N}(x) & = & 0, & \quad x\le -1 \medskip \\
Z_{N=0}(x) & = & \delta_{x,0},
\end{array} \right.
\label{eq:28}
\ee
Results of direct iterations of \eq{eq:28} presented in \fig{fig:05} show that depending on values
of vertex degrees, $p$ and $p_0$, the localization of the trajectories may or may not exist: one
clearly sees different behavior of $Z(N,x|p,p_0)$ as a function of $x$ for few values of $p_0$
above and below the transition point, $\bar{p}_0=p$.

As previously discussed in Section II, half of the values of the partition function $Z_{N}(x)$
(those corresponding to odd values of $(N+x)$) equal zero. Therefore, without loss of information
one can replace $Z_{N}(x)$ by $V_N(y)$ defined as follows:
\be
\left\{\begin{array}{rcll} V_{N}(y) & = & \disp Z_N(2y-1)+Z_N(2y), & k\ge 1 \medskip \\
V_{N}(0) & = & Z_N(0) \\
\end{array} \right.
\ee
This new partition function $V_N(k)$ satisfies
\be
\left\{\begin{array}{rcll} \disp V_{N+1}(k) & = & V_N( k-1) + (p-2)V_N(k) + V_N(k+1), &  \quad k>1
\medskip \\ \disp V_{N+1}(k) & = & p_0 V_N(k-1) + (p-2)V_N(k)+V_N(k-1), & \quad k=1 \medskip \\
V_{N+1}(k) & = & V_N(k+1), & \quad k=0 \medskip \\
V_{N}(k) & = & 0, & \quad k\le -1 \medskip \\
V_{N=0}(k) & = & \delta_{k,0},
\end{array} \right.
\label{eq:29}
\ee
It turns out that this set of equations is nothing but the set of equations describing the path
counting problem on regular hyperbolic graphs defined above for a particular case of $b=p-2$.

Indeed, writing down explicitly the recursion relation for a regular hyperbolic graph one gets:
\be
\left\{\begin{array}{rcll} \disp Z_{N+1}(x) & = & (p-b-1)Z_N( x-1) + bZ_N(x) + Z_N(x+1), &  \quad
x>1
\medskip \\ \disp Z_{N+1}(x) & = & p_0 Z_N(x-1) + bZ_N(x)+Z_N(x-1), & \quad x=1 \medskip \\
Z_{N+1}(x) & = & Z_N(x+1), & \quad x=0 \medskip \\
Z_{N}(x) & = & 0, & \quad x\le -1 \medskip \\
Z_{N=0}(x) & = & \delta_{x,0},
\end{array} \right.
\label{eq:30}
\ee
In what follows we solve the more general case of Eq.\eq{eq:30}, and then obtain the result for
decorated star graphs substituting $b=p-2$ in the final expression. The solution below is
completely analogous to what is presented above in Section III. Making a shift $x\to x+1$ and
symmetrizing \eq{eq:30} by substitution:
\be
Z_N(x)=(p-b-1)^{N/2}(p-b-1)^{x/2} W_N(x),
\label{eq:AB}
\ee
results in
\be
\left\{\begin{array}{rlll} W_{N+1}(x) & = & \disp W_N(x-1) + W_N(x+1) +
\frac{b}{\sqrt{p-b-1}}W_N(x) + \\ & + &\disp \frac{p_0-(p-b-1)}{p-b-1}\,\delta_{x,2}\,W_N(x-1)-
\frac{b}{\sqrt{p-b-1}}\,\delta_{x,1}\,W_N(x), & \qquad x\ge 1 \medskip \\
W_N(x) & = & 0, & \qquad x=0 \medskip \\
W_{N=0}(x) & = & \disp \frac{\delta_{x,1}}{\sqrt{p-b-1}}
\end{array} \right.
\label{eq:31}
\ee
Performing the $\sin$-Fourier transform for the generating function (similarly to
\eq{eq:07}--\eq{eq:08}) we obtain an integral equation
\be
\frac{\tilde{{\cal  W}}(s,q)}{s} - \disp \frac{\sin q}{s\sqrt{p-b-1}} = (2\cos q+B)\, \tilde{{\cal
W}}(s,q) +\frac{2}{\pi} \left[  A \sin 2q -B\sin q \right] \int_{0}^{\pi} \tilde{{\cal W}}(s,q)
\sin q\, dq
\label{eq:32}
\ee
where
\be
\tilde{{\cal  W}}(s,q) = \sum_{x=0}^{\infty}{\cal W}(s,x) \sin qx, \qquad {\cal  W}(s,x) =
\sum_{N=0}^{\infty}W_N(x) s^N
\ee
and
\be
A=\frac{p_0-(p-b-1)}{p-b-1}, \qquad B=\frac{b}{\sqrt{p-b-1}}.
\ee

Expressing $\tilde{\cal W}$ from \eq{eq:32} and integrating over $q$, $q\in [0,\pi]$ with the
weight $\sin q$, we obtain
\begin{multline}
I(s) = \disp \int_0^{\pi} \tilde{{\cal W}}(s,q) \sin q \, dq = \frac{1}{\sqrt{p-b-1}} \int_0^{\pi}
\frac{\sin^2q}{1-2s\cos q-Bs}\, dq \\ + \frac{2s}{\pi} \, I(s) \, \left[A\int_0^{\pi} \frac{\sin q
\sin 2q}{1-2s \cos q-Bs}\, dq - B\int_0^{\pi} \frac{\sin^2 q }{1-2s \cos q-Bs}\, dq \right]
\label{eq:33}
\end{multline}
The solution of \eq{eq:33} reads
\be
I(s) =\frac{\pi \sqrt{p-b-1}\left(1-Bs-\sqrt{(1-Bs)^2-4s^2} \right)}{4s^2-\disp
A\left(1-Bs-\sqrt{(1-Bs)^2-4s^2} \right)^2+2Bs\left(1-Bs-\sqrt{(1-Bs)^2-4s^2} \right)}
\label{eq:IS}
\ee
Performing the inverse Fourier transform, one gets an explicit expression for the generating
function ${\cal W} (s,x)$:
\begin{multline}
{\cal W}(s,x)=\frac{2}{\pi}\int_0^{\pi}\tilde{{\cal W}}(s,q) \sin qx\, dq \\ =
\frac{1}{s^2\sqrt{p-b-1}}\left(\frac{1-Bs-\sqrt{(1-Bs)^2-4s^2}}{2s}\right)^x
\left[1+\frac{2}{\pi}\sqrt{p-b-1}\, I(s) (2A-Bs)\right]
\label{eq:40}
\end{multline}

Finally, proceeding in a way analogous to the one in Section III, we obtain an explicit expression
for the generating function $Z(\sigma|p,p_0)$ of all paths of length $N$:
\begin{multline}
\bar{{\cal Z}}(\sigma|p,p_0) \equiv \sum _{x=0}^\infty  (p-b-1)^{x/2} {\cal W}(\sigma
\sqrt{p-b-1},x)= \\
= \frac{1}{\sigma (p-b-1)}\frac{1}{2\sigma-1+b\sigma+\sqrt{(1-b\sigma)^2-4\sigma^2(p-b-1)}}\times
\medskip \\ \times \left[ 1+\disp
\frac{2}{\pi}I(\sigma\sqrt{p-b-1})\left(2\,\frac{p_0-(p-b-1)}{p-b-1}-b\right)\right].
\label{eq:41}
\end{multline}

This function, once again has three singularities: the branching point of the square root
\eq{eq:41}:
\be
\sigma_1 = \frac{1}{b+2\sqrt{p-b-1}}.
\label{eq:sing1_1}
\ee
corresponding to the border of the continuous spectrum, the zero of the denominator of \eq{eq:41}
\be
\sigma_2 = \frac{1}{p}
\label{eq:sing2_2}
\ee
corresponding to the spreading wave solution, and the third singularity, $\sigma_3$, corresponding
to the zero of the denominator of  $I(\sigma\sqrt{p-b-1})$ (see \eq{eq:IS}), which correspond to
the localized solution. Substituting $\sigma_2$ into equation for $\sigma_3$ provides the condition
for critical value of $p_0$ which separates localized and delocalized regimes for regular
hyperbolic graphs:
\be
\bar{p}_0=p(p-b-1),
\label{eq:cr}
\ee
for decorated star graphs $b=p-2$ and this condition reduces to simple $\bar{p}_0=p$ in full
agreement with numerical simulations presented above, while for a tree-like graph ($G$) with $b=0$,
\eq{eq:cr} is reduced to \eq{eq:18}.

It seems interesting to compare the statistics of trajectories on the simplest star-like graph $G$,
shown in the \fig{fig:01}a and on the decorated one, $G_d$, depicted in the \fig{fig:04}a. The
mean-square displacement of the end of $N$-step path on a \emph{single branch} of $G$ and of $G_d$,
scales as $\sqrt{N}$ in both cases (with different numeric coefficients). Besides, the trajectories
on the graph $G$ are localized for $p_0\ge 3$, and on the decorated graph $G_d$ the localization
occurs at $p_0\ge 4$ (for $p_0=3$ the paths on $G_d$ are delocalized).

\section{Discussion}

In this paper we study localization properties for a path counting problem on several classes of
regular graphs with a single special vertex (trees with a "heavy" root, decorated stars and regular
hyperbolic graphs). Generalizing the argument of \cite{ternovsky} we show that in all those cases
the special vertex with functionality larger than that of the regular ones, works as an entropic
trap for the paths on the graph, and may lead, if the trap is strong enough, to a \emph{path
localization}.

We used two different techniques: the study of the spectral properties of the graph adjacency
matrix for finite graphs, and the study of singularities of the grand canonical partition function
in case of infinite graphs. In our opinion, parallel consideration of these two approaches has in
itself a significant methodical value, allowing the reader to see the similarity of these methods,
and the ways how the same values can be interpreted in two different languages.

Despite, to the best of our knowledge the results presented in this paper add some new flavor to
path localization on inhomogeneous graphs and networks, they certainly are an addition to the
long-standing theory of localization in disordered systems, whose development goes back all the way
to the works of I.M Lifhsitz \cite{lifshitz,pastur}.

There is a variety of problems in physics of disordered systems and inhomogeneous media whose
solutions (e.g., solutions of corresponding hyperbolic or parabolic equations, leading eigenvectors
of corresponding operators, etc.) are localized in the vicinity of some spatial regions. Similar
localization problems are often studied in polymer physics, where they correspond to a polymer
chain being adsorbed at some specific locations in space: like point-like defects of texture,
within some particular region of space, or in the vicinity of interfaces. Most commonly, the reason
for such localization is energetic: it is due to some attractive force between localized particle
or polymer chain and the absorbing substrate.

The situation described in this paper belongs to a class of problems for which the origin of
localization is purely \emph{entropic} and is caused exclusively by geometric reasons. Among
similar problems discussed in the literature is, first of all, the localization of ideal polymer
chains on regular lattices with defects discussed in \cite{burda} and trapping of random walks in
inhomogenious media \cite{balagurov,donsker}. Similar situations are also widely discussed in
spectral geometry, where the solutions of Laplace or Helmholtz equations in regions of complex
shape (e.g., obtained by gluing together several simple shapes) are studied. The principal question
there consists in determining the condition on the localization of wavefunctions in these complex
regions \cite{grebenkov}.

The phase transition on regular graphs, though in slightly different probabilistic setting, has
been discussed recently in \cite{vershik2}, where authors exhaustively described the exit boundary
of random walks on homogeneous trees. They showed that the model exhibits a phase
transition, manifested in the loss of ergodicity of a family of Markov measures, as a function of
the parameter, which controls the local transition probabilities on the tree. We plan to analyze
whether the transition found in \cite{vershik2} has direct relations to the transition described in
our work.

One last remark we would like to make concerning the results obtained above, is as follows. It is
seen from our results that the condition for localization to occur depends significantly on the
overall geometry of the graph. Indeed, on decorated star graphs it is enough to have $p_0>p$ to
have a localization, while on tree-like graphs one needs $p_0>p(p-1)\sim p^2$ in order for
trajectories to be localized. Accordingly, it remains unclear if it is possible, by changing the
graph geometry, to push the transition value $\bar{p}_0$ below $p$, so that localization will occur
even on a graph with all vertices having the same degree $p$. One possible candidate for such a
localization might be a regular random graph (i.e., a random graph whose all vertices have the same
degree $p$): by chance on such graphs there is a small (of order one) number of short cycles. In
principle it might be conceivable that these cycles work as entropic traps in a way similar as
discussed in this paper. The localization on a family of specially prepared random regular graphs
with enriched fraction of short loops has been recently studied in \cite{valba1,valba2}.

\begin{acknowledgments}
The authors are grateful to V Avetisov, Z Burda, V Chernyshev, D Grebenkov, A Gorsky,  V Kovaleva,
and A Vershik for valuable discussions as well as to A. Maritan for pointing us the Ref [17]. This work was partially supported by RFBR grant no. 16-02-00252 and by EU-Horizon 2020 IRSES project DIONICOS (612707). O.V. acknowledges support of the Higher School of Economics program for Basic Research.
\end{acknowledgments}

\small{\emph{Note added in proof.} Z. Burda pointed out that in \cite{18} the system very similar to
the one described in our Section 2, has been discussed. However, the authors of \cite{18}
considered finite tree-like graphs where there is no localization transition.}

\end{document}